# Coherent optical modulation of partially mode-locked fiber laser based on coherent population oscillation in reduced oxide graphene

Lei Gao[1,2], Yulong Cao[1,2], Hongqing Ran[1], Lingdi Kong[1], Yujia Li[1], Ligang Huang[1], Wei Huang[1], Danqi Feng[1], and Tao Zhu[1]

*1 Key Laboratory of Optoelectronic Technology & Systems (Ministry of Education), Chongqing University, Chongqing 400044, China*
*2 These authors contributed equally to this work*
*Corresponding author: gaolei@cqu.edu.cn; zhutao@cqu.edu.cn*

**Optical control of graphene-based photonic devices and systems has been under extensive explorations, nevertheless, the requirement of high power pump laser due to incoherent modulation makes those schemes low efficient. Here, we demonstrate coherent manipulation of the operating states of partially mode-locked fiber laser based on coherent population oscillation in reduced graphene oxide for the first time. We couple a much weaker continuous wave laser into the resonator operating with parametric instability state, and observe significant depression/enhancement of the sidebands when the coherent population oscillation conditions are satisfied. Besides, significant depression of partially mode-locked fiber laser is achieved. The experimental results reveal that the coherent population oscillation in reduced graphene oxide is highly effective in manipulating mode-locked fiber laser system, and the induced phase variation is highly asymmetrical. The discovery of coherent population oscillation in reduced graphene oxide facilitates coherent optical modulation of reduced graphene oxide-based photonic devices and systems with a much weaker controlling laser.**

Active control of laser systems through tunable devices has been under extensive explorations for critical importance in both fundamental physics and practical applications [1-4]. Nevertheless, conventional systems based on mechanical deformation methods are unstable. As a two-dimensional Dirac material with zero bandgap, graphene proves to be an ideal choice for reconstructing controllable optical devices for its unique characteristics, including controllable carrier dynamics together with ultrafast response, ultrahigh electron mobility, and low linear absorption over a wide spectral range [5,6]. Through tuning the filter bandwidth, phase, modulation depth, and dispersion of graphene-based devices, a variety of works have been explored, based on optical thermal effect, external electrical biasing, and optical modulation [7-16]. The electric field-control capability motivates optoelectronic devices with extraordinary performances, while the modulation speed is limited by the bandwidth of bias circuit. In contrast, optical modulation of graphene-based devices makes the tuning process rather stable, and also breaks the "electrical bottleneck" in electric-controlled method, extending the modulation bandwidth from ~1 GHz to ~200 GHz and even beyond [10-13].

When graphene is excited by a strong pump laser with photon energy near the Dirac-zone, electrons are excited from valence band to conduction band, then the formed Fermi-Dirac distribution blocks absorption of a weak probe laser due to Pauli blocking. The optical modulation of graphene expedites diverse optically controlled devices and systems, such as switcher, and modulator [7,13]. Sheng et al tune the pulse duration of a soliton laser from 705 fs to 1356 fs through optical modulation depth of the saturable absorber with external laser [7]. We also report the pulse duration regulation through tuning the effective bandwidth of graphene-covered-chirped fiber Bragg grating by changing refractive index of graphene [17]. Nevertheless, this process is incoherent, which is mainly originated from the carrier-phonon scattering process of graphene [18]. Therefore, the strong power of pump laser makes the incoherent optical modulation and thermal-optical effects low efficient. It is necessary to control the phase of graphene with a much weaker pump laser.

Recently, we demonstrate coherent optical modulation of graphene based on coherent population oscillation (CPO), where population of the carriers oscillate with a beat frequency determined by the pump and probe frequency difference [19]. Then, the absorption of the probe laser is reduced with a frequency width on the order of $1/2\pi T_2$ [20-22], where a picosecond scale $T_2$ is the relaxation time associated with electron interband relaxation and cooling of hot phonons. Due to the Kramers-Kronig relations, the phase of the probe laser is modulated coherently by the pump laser through CPO. Depending on fabrication process, number of layers, the detected value $T_2$ are different [23, 24]. We observe a burning hole width of 34.8 GHz for single layer graphene produced by chemical vapor deposition [19]. Simultaneously, as another important way to produce graphene, the reduced graphene oxide (rGO), which is fabricated by reducing graphene oxide via chemical, thermal or electrochemical means, behaves similar optoelectronics response as pristine graphene [25-27]. The interband carrier recombination times $T_2$ after excitation are different due to energy band gaps related to the size of the nanometer-scale $sp^2$ clusters in various rGO materials. Here, we take a typical $T_2$ of 4.3 ps [27], and the expected spectral width of giant phase changing related to the CPO in rGO is 36.9 GHz. More potential devices and systems are possible via tuning the phase of rGO through CPO.

In this letter, we coherently manipulate the operating states of partially mode-locked fiber laser (PML) based on CPO in rGO for the first time, with a much weaker pump laser compared with that in incoherent optical modulation. We tune the polarization states of the laser system, and obtain a high coherent state dominated by parametric instability (PI), together with deteriorated coherent state of partially mode-locking state due to vector stochastic four-wave-mixing (FWM) among PI gain bands. By coupling a continuous wave laser (CWL) with controllable state of polarization (SOP) into the resonator, significant depression/enhancement of the PI gain bands are obtained. Besides, depression of PML is observed when the CPO condition is satisfied.

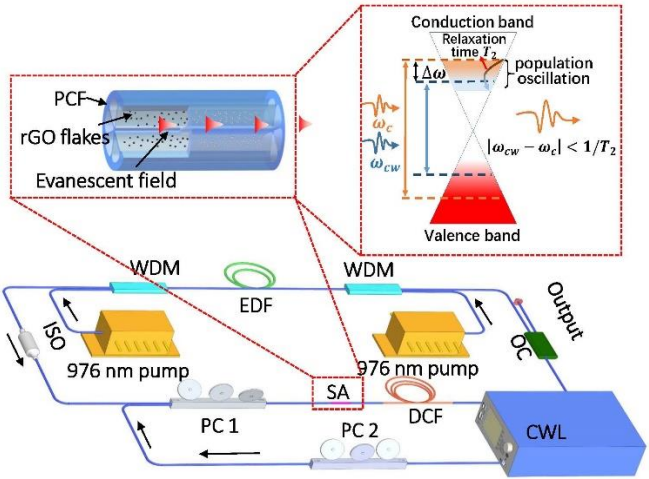

Fig. 1. Setup of the coherent optical modulation of partially mode-locked fiber laser. The inset illustrates CPO effect in graphene with frequency detuning smaller than the inverse of the relaxation time $T_2$.

The main setup is shown schematically in Fig. 1. The ring fiber cavity consists 1 m erbium-doped fiber (EDF), a wavelength division multiplexer (WDM), a polarization independent optical isolator (ISO), a polarization controller (PC), 19.5 m dispersion compensation fiber (DCF), an optical coupler (OC), and a saturable absorber (SA) made by filling reduced graphene oxide (rGO) flakes into cladding holes of a photonic crystal fiber (PCF) [28]. The system is used previously to generate PML, and its formation process based on vector FWM among parametric gain lobes has been explained in detail in Ref. 28. A well fit of Lorentzian shape of 2D band in the Raman spectrum implies that the graphene flakes are electronically decoupled, and they maintain Dirac fermions linear dispersion. Here, we couple an external CWL with controllable SOP, wavelength, and power, into the cavity with a counterclockwise direction. The external CWL circulates in the cavity, and it passes through the SA many times. As the laser circulating in main cavity possess the same direction with the external CWL, we can monitor the SOPs of the two lasers simultaneously. The line width of the external CWL is 0.05 nm, and its power is much smaller than that of the system, hence, its influence on the gain property of the EDF can be ignored. Therefore, the different operating states of the laser system are mainly result from the variation of phase-matching condition due to the nonlinear phase shift of rGO under illumination of the external CWL.

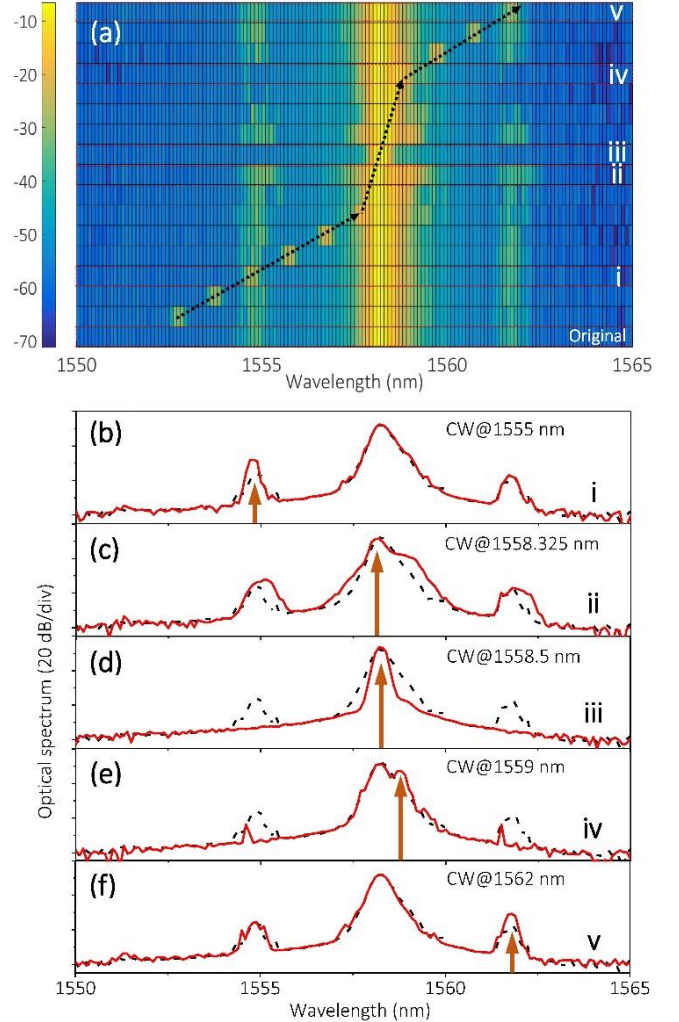

**Fig. 2.** (a) Optical spectra of the PI under different CWL wavelengths, the intensities are shown with dB scale. The dotted line presents the center wavelength of the CWL from blue side to the red side with two step sizes (Far from center region: 1 nm/step; center region: 0.175 nm/step). (b)-(f) Spliced optical spectra from (a) with respect to the original spectrum. The dotted line is the original spectrum of PI.

With proper cavity detuning, this system may generate primary gain lobes alone, namely, the excitation of parametric instability (PI). Physically, it is a degenerate FWM, and the quasi-phase-matching condition can be described as $2\gamma P + \beta_2 \Delta\omega_k^2 = 2k\pi/Z$, where, $\gamma$, $P$, $\beta_2$, $Z$ are nonlinear coefficient, peak power, average group velocity dispersion, and dispersion period, respectively [28, 29]. $\Delta\omega$ is the frequency shift from the center, and $k$ presents the sidebands order. PI occurs when the phase accumulated in each round trip equals to

2kπ. The sidebands can be enhanced and depressed when the cavity is off resonant through changing the phase by biasing PC, or modulating rGO by external CWL through CPO effect.

Primarily, we set two 976 nm lasers to 400 mW, and the power of the external CWL as zero. PI gain sidebands are obtained when rotating PC properly. The optical spectrum of the PI shown in Fig. 2 is denoted as original. The two sidebands generated by degenerate FWM are relatively weak with respect to the center wavelength, and SOP of the output is fixed as a closely clustered spot on the Poincaré sphere. The coherence of the PI is high. The wavelength of the external laser with a power of 1 mW is fixed at 1558.5 nm, which is the center wavelength of PI. The SOP of the CWL is tuned to cover the whole space of the Poincaré sphere, and significant changes occur only when the SOP of the external laser is located on the same spot of PI in the Poincaré sphere. For CWL at this wavelength, depression of sidebands can be observed for average power larger than 0.5 mW, which is much smaller than that in Refs. 7, 17, and 18.

The influence of the lasing wavelength of the external laser on PI is investigated. Here, the average power is fixed at 1 mW, and the SOP is overlapped with that of PI. The optical spectra of the PI under different CWL wavelengths are presented in Figs. 2 (b)-(f). When the wavelength of external laser is scanning from 1553 nm to 1562 nm, significant changes can be observed only near the center wavelength region. It should be noticed that the enhancement of PI occurs on the blue side of the center wavelength (in Fig 2. (c)) within a wavelength detuning of ~0.175 nm, while great depression of PI appears on the center wavelength (Fig. 2 (d)). This depression effect diminishes gradually when the external laser is tuned far away from the center wavelength in the red side with a wavelength detuning of ~0.5 nm.

A more important thing we care is about the required power of the external CWL for manipulating the laser states. The average power of the CWL for depressing the PI sidebands here is less than 0.5 mW, when the SOPs of the two lasers are overlapping. Yet, a average power of the CWL larger than 50 mW is needed when the condition of CPO effect is no longer satisfied. For the main laser system described above, diverse operating states may be generated based on various nonlinear phase-matching conditions. For specific biasing of PC1, we obtain PML with broadband optical spectrum, where cascaded vector FWM occurs among PI sidebands. When the cavity gain is high enough, newly generated frequencies merge into a gap-free comb. The longitudinal modes are populated with a random distribution of position, intensity and polarization. It has also been referred as noise-like pulses: bunches of pulses with irregular varying time duration and intensity. The autocorrelation trace of the PML contains a coherence peak of femtosecond duration, sitting on a broad pedestal with a duration of picoseconds. The details are shown in Ref. 28. Here, we perform the same procedures as in the manipulation of PI process, and similar results are obtained. We plot the SOPs for filtered output at 1556.1 nm, namely the first anti-Stokes gain band in Fig. 3 (a). We find that the polarization states for the PML and pure PI are highly different. The detected polarization state of PI is a fixed point in the Poincaré sphere, while it bifurcates to randomization for the system operating with PML, which is described in our previous work [28], defined as a SOP pool with random intensities.

Similar to the procedures in the depression of PI in PML, we adjust the SOP of the external CWL via biasing the PC2 from far away to the SOP pool of the main laser system. The center wavelength of the CWL is 1558.5 nm, which is the center wavelength of PML, and the average power is 50 mW. As shown in Fig. 3 (b), depression of PML occurs only when the SOP of the external CWL locates in the SOP pool of the laser output. This phenomenon is same as the depression of PI in PML, in that depression occurs only when the SOP of external CWL is overlapping with that of the main laser output. The SOP tolerance seems loosen as the SOP of the main laser system is a broad pool rather than a fixed spot for pure PI. Besides, we find that the average power of the external CWL (~35 mW) for significant depression of PML is much higher than that of depression of PI (~0.5 mW). The outputs under different CWL powers are shown in Fig. 3 (c). The increment of the threshold is understandable as a much higher power is needed for FWM between CWL and PML. The depression of PML also requires specific frequency detuning between CWL and PML, just as that in the depression of PI. Figure 3 (d) depicts the outputs for various powers of CWL. The center wavelength of the CWL is the same of that as the first anti-Stokes. No significant depression of PML has been found even the average power is larger than 80 mW.

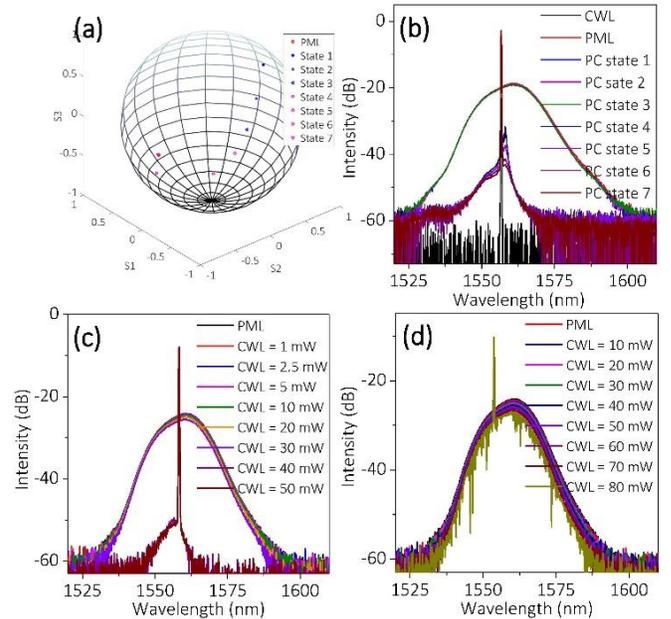

Figure 3. (a) Center of the SOP pool of the PML and the various SOPs of the CWL, and (b) corresponding spectral outputs. (c) Outputs for CWL with different powers. The wavelength of CWL is 1558.5 nm, which is also the center wavelength of PML. (d) Outputs for CWL with different powers, the wavelength of CWL is 1556.1 nm, which is the center wavelength of the first anti-Stokes sideband.

Considering the relatively narrow detuning frequency region (0.765 nm), the contribution from gain competition and the saturable absorption effect from EDF can be neglected reasonably. Besides, the incoherent optical modulation effect can also be excluded as the modification of PI and PML disappears when the SOP of the CWL is different from that of the main system, as the external CWL modifies CPO only when they are in-phase. The enhancement and depression of the PI sidebands are originated from the variation of the quasi-phase-matching conditions through modulating the nonlinear phase via CPO in rGO with the external CWL. As the PI in this laser system is highly coherent, the phase experienced by the PI process is a straight evidence of the CPO induced by the laser and the CWL. Besides, the variation in the spectral width of ~0.765 nm (84.4 GHz in in C band), which is comparable to the predicted value of 36.9 GHz. The possible reason for the discrepancy is that the relaxation time $T_2$ in our experiment may be different from that in Ref. 27. More interestingly, we found that the phase variation exhibits giant asymmetry. Similar asymmetry has been observed in CPO in single layer graphene [19]. They share the same physics, such as FWM between the center wavelength and the CWL. Yet, the even sharp asymmetry in CPO may originate from the disorder in rGO. Besides, the variation of the phase difference, $\theta$, between the center wavelength and the CWL may also lead to asymmetry of the burning hole of CPO [22,23]. When $\theta=0$, the depth of burning hole reaches to maximum due to the constructive interference of pure CPO and FWM, while the burn hole disappears for destructive interference with $\theta=\pi$. When $0<\theta<\pi$, for example $\theta=\pi/2$, both the dispersion and absorption lines show significant asymmetry, and Fano shape describing quantum interference of photons and electron-hole pairs appears in the absorption curve [20,22]. That's to say, the phase variation of rGO can be manipulated either positively or negatively by this phase detuning.

In summary, we have demonstrated coherently optical modulation of the states of partially mode-locked fiber laser system based on CPO in rGO. We find that on the coherent conditions, the required power of pump laser is much smaller than that in incoherent optical modulation schemes. The degenerate FWM of external CWL and center laser modifies the CPO effect, leading to an asymmetrical burning hole. Such an intense variation of dispersion is utilized to changes the phase-matching condition of PI and PML, and asymmetrical enhancement and depression of the operating states are experimentally observed. The CPO in rGO circumvents the "electrical bottleneck" in electrical field tuning of Fermi level, and require a much weaker pump power for manipulating rGO-based optoelectronic devices and systems. Such a kind of high coherent optical modulation will motivate diverse graphene-based controllable photonic devices with extraordinary performances.

**Funding.** Natural Science Foundation of China (61635004, 61405023), the National Postdoctoral Program for Innovative Talents (BX201600200), the Postdoctoral Science Foundation of China (2017M610589), the Postdoctoral Science Foundation of Chongqing (Xm2017047), the Science Foundation of Chongqing (CSTC2017JCYJA0651), the Science and Technology on Plasma Physics Laboratory (6142A0403050817), the National Science Fund for Distinguished Young Scholars (61825501).